\documentclass[%
 reprint,
showpacs,
 amsmath,amssymb,
 aps,
 pra,
longbibliography,
floatfix
]{revtex4-2}

\usepackage[utf8]{inputenc}	
\usepackage[T1]{fontenc}	
\usepackage[]{graphicx}		
\usepackage{xcolor}
\usepackage{gensymb}
\usepackage{braket}
\usepackage{comment}
\usepackage[english]{babel}

\renewcommand{\b}{\hat{b}}
\renewcommand{\c}{\hat{c}}
\renewcommand{\a}{\hat{a}}
\newcommand{\U}{\hat{U}}
\newcommand{\I}{\hat{I}}
\renewcommand{\L}{\hat{L}}

\renewcommand{\H}{\hat{H}}

\begin{document}

\title{Interaction of quantum systems with single pulses of quantized radiation}

\author{Victor Rueskov Christiansen}
\email{victorrc@au.dk}
\affiliation{
Center for Complex Quantum Systems, Department of Physics and Astronomy,
Aarhus University, Ny Munkegade 120, DK-8000 Aarhus C, Denmark}

\author{Alexander Holm Kiilerich}
\affiliation{
Numerical Competence Centre,
Ørsted A/S, \\ Kraftværksvej 53, Skærbæk, DK-7000 Fredericia, Denmark}

\author{Klaus M{\o}lmer}
\email{moelmer@phys.au.dk}
\affiliation{Aarhus Institute of Advanced Studies, Aarhus University, H{\o}egh-Guldbergs
Gade 6B, DK-8000 Aarhus C, Denmark}
\affiliation{
Center for Complex Quantum Systems, Department of Physics and Astronomy,
Aarhus University, Ny Munkegade 120, DK-8000 Aarhus C, Denmark}

\date{\today}

\bigskip

\begin{abstract}
The interaction of a propagating pulse of quantum radiation with a localized quantum system can be described by a cascaded master equation with a distinct initially populated input and a finally populated output field mode [Phys. Rev. Lett.{\bf 123}, 123604 (2019)]. 
By transformation to an appropriate interaction picture, we identify the usual Jaynes-Cummings Hamiltonian between the scatterer and a superposition of the initial and final mode, with a strength given by the travelling pulse mode amplitude. The transformation also identifies a coupling of the scatterer with an orthogonal combination of the two modes. The transformed master equation offers important insights into the system dynamics and it permits numerically efficient  solutions.  
\end{abstract}

\maketitle
\noindent

\section{Introduction}

A full quantum analysis of the interaction of a discrete quantum system with a cavity field may often assume (near) resonance with only a single standing wave field mode whose mode function is unaffected by the interaction. The interaction then only implies the joint evolution of a  single  field oscillator and the discrete quantum system, as described, e.g., in the Jaynes-Cummings model. 


Numerous proposals exist to employ interactions with simple quantum systems with the purpose to manipulate and prepare quantum states, such as number states, coherent states, squeezed states and Schr\"odinger cat states of a single mode of radiation. The motivation is often that such states can be propagated in space and thus be used for probing, communication, and transfer of states and operations in quantum networks \cite{kimble2008quantum}. But, how does a localized quantum system, such as a single two-level atom, interact with a travelling pulse that initially occupies only a single mode of radiation?  If the temporal profile of this pulse at the location of the atom is $u(t)$, do we obtain a time-dependent interaction of the same form as for an atom flying through a cavity and exploring the position dependent field strength of the resonant eigenmode? That is, is the system correctly or to a good approximation described by the Hamiltonian $(\hbar=1)$,
\begin{equation}\label{eq:JCMpulse}
H_{JC}(t)=i\sqrt{\gamma}u(t)(\a^\dagger \sigma^- - \a \sigma^+),
\end{equation} 
where $\sqrt{\gamma}u(t)$ specifies the time dependent coupling strength, and the raising and lowering operators of the field and the two level system account for the coherent exchange by absorption and emission of quanta of excitation between the atom and the travelling pulse?

The answer to this question is complicated by the fact that in the absence of a cavity, the propagating field explores a continuum of frequency modes and by dispersive and absorptive effects, the interaction with a scatterer may change the temporal shape of the field mode function in a manner that is entangled with its quantum state of excitation. Multiple analyses have addressed different aspects of this multi-photon, multi-mode problem \cite{gheri1998photon,PhysRevA.92.033803,PhysRevA.86.013811,PhysRevA.86.043819,PhysRevA.96.023819,PhysRevX.7.041010, zhang2021control, Shen:05, Fischer2018scatteringintoone, shi2015}. 

As a simpler problem, we may enquire what is the quantum state occupying a definite pulse mode after the interaction. Applying the theory of cascaded quantum system \cite{gardiner,Carmichael1993}, that problem was treated in Refs.\cite{short_kiilerich,long_kiilerich} by incorporating the incident pulse of quantized radiation as the output from an upstream virtual single mode cavity that gradually releases its quantum state content in the form of a pulse. The quantum state of any specific output pulse mode may, in a similar manner, be associated with the asymptotic final content of downstream filter cavity. This experimentally inspired construction  of input and output wave packet modes leads to a simple density matrix formalism with time dependent couplings of the scatterer to two discrete cavity modes. The problem thus takes a quite different form than suggested by  Eq.\eqref{eq:JCMpulse}.   


The cascaded master equation describes dynamics where the upstream cavity (representing the incident pulse) leaks all its quanta, while the downstream cavity gradually acquires the entire fraction of the output field that populates the specified output pulse mode function.
In this article we derive the master equation in the interaction picture with respect to the undisturbed propagation of the pulse, represented by a linear transfer of quanta between the upstream and downstream cavities. In this interaction picture, the coupling of the field to the scatterer takes a simple form and numerical solutions are much less demanding.

The article is structured as follows: In section \ref{sec:virtual_cavity_approach} the virtual cavitites and cascaded system formalism is presented. In section \ref{sec:interaction_picture} we introduce the interaction picture and the corresponding master equation for the case of identical input and output modes. In section \ref{sec:three_mode_interaction_picture} we extend the interaction picture to describe different input and output pulses, which includes for example the effect of dispersion caused by an empty cavity. As a numerical example we  analyze the creation of squeezed states and Schr\"odinger cat states by the scattering of pulses on a cavity with a Kerr non-linearity.

\section{Quantum interactions with a light pulse - a virtual cavity approach} \label{sec:virtual_cavity_approach}

Consider a local quantum system described by the Hamiltonian $\H_{\textup{s}}(t)$ and Lindblad dissipation terms $\L_i$, with $i=1, ... n$. Its interaction with an incident quantized radiation field is governed by  $V=i\sqrt{\gamma}[\c \b_{in}(t)^\dagger - \c^\dagger \b_{in}(t)]$, where $\c^{(\dagger)}$ is the lowering (raising) operator of the system accompanying absorption (emission) of a quantum of radiation. This interaction involves the input field continuum operators $\b_{in}(t)$ which can be viewed as the Fourier transform of the frequency eigenmode annihiliation operators. If the incident radiation is restricted to a single pulse, described by a square integrable mode function $u(t)$, the interaction can be accounted for by an effective cascaded system master equation \cite{short_kiilerich}. Here the quantum pulse is described as if it leaks from an upstream virtual cavity with a coherent out-coupling strength $g_u(t)$, where
\begin{align}\label{eq:g1}
g_u(t)  = \frac{u^*(t)}{\sqrt{1-\int_0^t dt'\, |u(t')|^2}}.
\end{align}
Likewise the component of outgoing radiation that eventually occupies an arbitrary wave packet mode  $v(t)$, can be picked up by a virtual downstream filter cavity with a coherent in-coupling strength $g_v(t)$,
\begin{align}\label{eq:g2}
g_v(t) = -\frac{v^*(t)}{\sqrt{\int_0^t dt'\, |v(t')|^2}}.
\end{align}

All other output modes are reflected by the $v$-cavity, and they are, in our formalism described as Markovian loss. According to input-output theory  \cite{gardiner}, the output field after reflection by the $v$-cavity is thus governed by the annihilation operator 
\begin{align}\label{eq:bout}
\b_{out}(t) = \b_{in}(t) + g_u^*(t)\a_u +\sqrt{\gamma}\c+g_v^*(t)\a_v.
\end{align}
which represents the interference between the amplitudes of the incident vacuum field operator and fields emitted by the scatterer and the two cavities. Detection of an outgoing photon is thus accompanied by the action of a single quantum jump operator, equivalent to the appearance of a Lindblad damping term, 
\begin{align}\label{eq:L}
\L_0(t) = \sqrt{\gamma}\c+g_u^*(t)\a_u+g_v^*(t)\a_v,
\end{align}
in the master equation for the joint state of the quantum scatterer and the two virtual cavities ($\hbar=1$),
\begin{align}
\frac{d\rho}{dt} = \frac{1}{i}[\H(t),\rho ]+\sum_{i=0}^n D[\L_i]\rho.
\end{align}
Here, the Hamiltonian is formed both by the system part $\H_{\mathrm{s}}(t)$ and the interactions between the different components,  
\begin{widetext}
\begin{align}\label{eq:H}
\begin{split}
&\H(t) = \H_{\mathrm{s}}(t)+\frac{i}{2}\big(\sqrt{\gamma}g_u(t)\a_u^\dagger\c
+ \sqrt{\gamma} g_v^*(t)\c^\dagger\a_v  + g_u(t) g_v^*(t)\a_u^\dagger\a_v - \mathrm{H.c.}\big)
\end{split}
\end{align}
\end{widetext}
and the master equation terms  $D[\L_i]\rho =  -\frac{1}{2}(\L_i^{\dagger}\L_i\rho + \rho\L_i^{\dagger}\L_i) + \L_i \rho \L_i^{\dagger}$ apply both for the outgoing field loss, represented by $\L_0$ in \eqref{eq:L} and for the damping terms $\{\L_{i=1,..n}\}$ acting on the quantum scatterer. Note that both the Hamiltonian and the $\L_0^{\dagger}\L_0$ product terms in the master equation give rise to  cross terms between the field and scatterer operators $\a_{u(v)}^{(\dagger)}$ and $\c^{(\dagger)}$, and the Hamiltonian and the dissipative terms conspire to cancel all contributions that cause excitation transfer towards the upstream cavity. This is a key property of the cascaded system master equation, built into its formal derivation \cite{gardiner,Carmichael1993}.

Reflection by a one-sided cavity is treated in a similar manner and it is also possible to treat multi-output situations where a scatterer causes both transmission and reflection  \cite{long_kiilerich}. If retardation effects can be neglected, more complex networks can also be treated   \cite{lodahl2017chiral}, while  non-Markovian effects of retardation and time delays engages the multi-mode character of the field \cite{zoller}.

\section{Time dependent modes - an interaction picture} \label{sec:interaction_picture}
The introduction of separate upstream and downstream cavity modes in our analysis enforces a two-mode treatment of the interaction between a single light pulse and a quantum system. Since the bare propagation of the light pulse without a scatterer amounts to a perfect release and recapture of the field by the input and output cavity modes, this suggest addressing the dynamics in an interaction picture where the transfer of the quantum state between these modes is already taken care of. The analysis of the problem gives exactly the same results, but the interaction picture should allow a much more efficient numerical treatment.

By the above argument, we hence pass to the interaction picture with respect to the cavity-cavity coupling component of \eqref{eq:H},
\begin{align}
    \hat{H}_0(t) = \frac{i}{2}[g_u(t)g_v^\ast(t)\hat{a}_u^\dagger\hat{a}_v
                             - g_v(t)g_u^\ast(t)\hat{a}_v^\dagger\hat{a}_u].
\end{align}

We thus define the unitary time evolution operator $\U_0(t)$ as the solution to,
\begin{align}\label{eq:U0}
i\frac{d}{dt} \U_0(t) = \H_0(t) \U_0(t)    \end{align}
with $\U_0(0)=\I$, the identity operator. The Schrödinger picture solution to the master equation can be written $\rho(t) = \U_0(t) \rho_I(t) \U_0(t)^{\dagger}$, where $\rho_I(t)$ solves the master equation in the interaction picture. In the interaction picture, $\H_0(t)$ is absent from the Hamiltonian, while the remaining terms and the Lindblad operators are transformed,  $\hat{O}_I(t) = \U_0(t)^{\dagger} \hat{O} \U_0(t)$.     

We recognize $\H_0(t)$ as a time dependent beam splitter-type coupling, and we make the ansatz
\begin{align}
    \hat{U}_0(t) = \exp(\lambda(t)\hat{a}_u^\dagger\hat{a}_v
                      - \lambda^\ast(t)\hat{a}_v^\dagger\hat{a}_u),
\end{align}
which upon insertion in Eq.\eqref{eq:U0} yields $\frac{d}{dt}\lambda(t) = \frac{1}{2}g_u(t)g_v^\ast(t)$. In the interaction picture, the cavity mode annihilation operators thus read
\begin{align} \label{eq:a_u_a_v_transformations}
\begin{split}
    \hat{a}_{u,I}(t) &= \cos\lambda(t)\hat{a}_u(0) + \sin\lambda(t)\hat{a}_v(0) \\
    \hat{a}_{v,I}(t) &= \cos\lambda(t)\hat{a}_v(0) - \sin\lambda(t)\hat{a}_u(0),
\end{split}
\end{align}
where $\hat{a}_{u(v)}(0)$ refer to the incident quantum field and the vacuum output field at the initial time, where the Schrödinger and the Interaction picture coincide. Note that the bare system Hamiltonian $\H_{\mathrm{s}}(t)$ and damping terms $\L_i^{\dagger}$, $\L_i$ with $i>0$ are not affected by the transformation to the interaction picture, which only concerns the pulse mode operators. We thus obtain
\begin{widetext}
\begin{align} \label{eq:arbitrary_hamiltonian}
    \hat{H}_I = \hat{H}_S + \frac{i\sqrt{\gamma}}{2} ((g_u(t)\cos\lambda^\ast+g_v(t)\sin\lambda^\ast)\hat{a}_u^\dagger \hat{c}
    + (g_v^\ast(t)\cos\lambda-g_u^\ast(t)\sin\lambda)\hat{c}^\dagger \hat{a}_v - \text{H.c}),
\end{align}
and
\begin{align} \label{eq:interaction_picture_lindblad}
    \hat{L}_{0, I} = \sqrt{\gamma}\hat{c} + g_u^\ast(t)(\cos\lambda\hat{a}_u + \sin\lambda\hat{a}_v) + g_v^\ast(t)(\cos\lambda\hat{a}_v - \sin\lambda\hat{a}_u)
\end{align}
\end{widetext}
In the following we shall omit the index $I$ with the understanding that all explicitly time dependent operators refer to the interaction picture.

\subsection{Identical incoming and outgoing modes}
We examine first the case where the outgoing field mode is equal to the incoming one, $v(t)=u(t)$, which suggests introducing
\begin{align} \label{eq:theta_definition}
    \sin^2\theta(t) \equiv \int_0^tdt'|u(t')|^2,
\end{align}
with $|u(t)|^2 = d\sin^2\theta/dt = 2\sin\theta\cos\theta d\theta/dt$, since this allows us to rewrite
\begin{align}
g_u^*(t) g_v(t) = \frac{-u(t)^2}{\sin\theta(t) \cos\theta(t)} =  -2 \frac{d\theta}{dt}.
\end{align}
We thus obtain the simple relation, $\lambda(t) = -\theta(t)$. When $u(t)$ is real, we can rewrite the interaction picture Hamiltonian  \eqref{eq:arbitrary_hamiltonian} as
\begin{widetext}
\begin{align} \label{eq:improved_interaction_hamiltonian}
    \hat{H}(t) = \hat{H}_S(t) + i\sqrt{\gamma}u(t)(\hat{a}_u^\dagger\hat{c} - \hat{c}^\dagger\hat{a}_u) 
    + \frac{i}{2}\sqrt{\gamma}u(t)(\cot\theta-\tan\theta)(\hat{a}_v^\dagger\hat{c} - \hat{c}^\dagger\hat{a}_v),
\end{align}
\end{widetext}
and the Lindblad operator \eqref{eq:interaction_picture_lindblad} of field loss as
\begin{align} \label{eq:improved_lindblad}
    \hat{L}_0(t) = \sqrt{\gamma}\hat{c} - (\tan\theta + \cot\theta)u(t)\hat{a}_v.
\end{align}
Remarkably, the first interaction term in Eq. \eqref{eq:improved_interaction_hamiltonian} is exactly the Jaynes-Cummings Hamiltonian \eqref{eq:JCMpulse} that one might have anticipated by simple arguments, but the interaction with the travelling pulse is supplemented by the coupling to a second mode of the field, and the dissipation of the system \eqref{eq:improved_lindblad} includes also this ancillary mode. 

\begin{figure}[b]
    \centering
    \includegraphics[width=\linewidth]{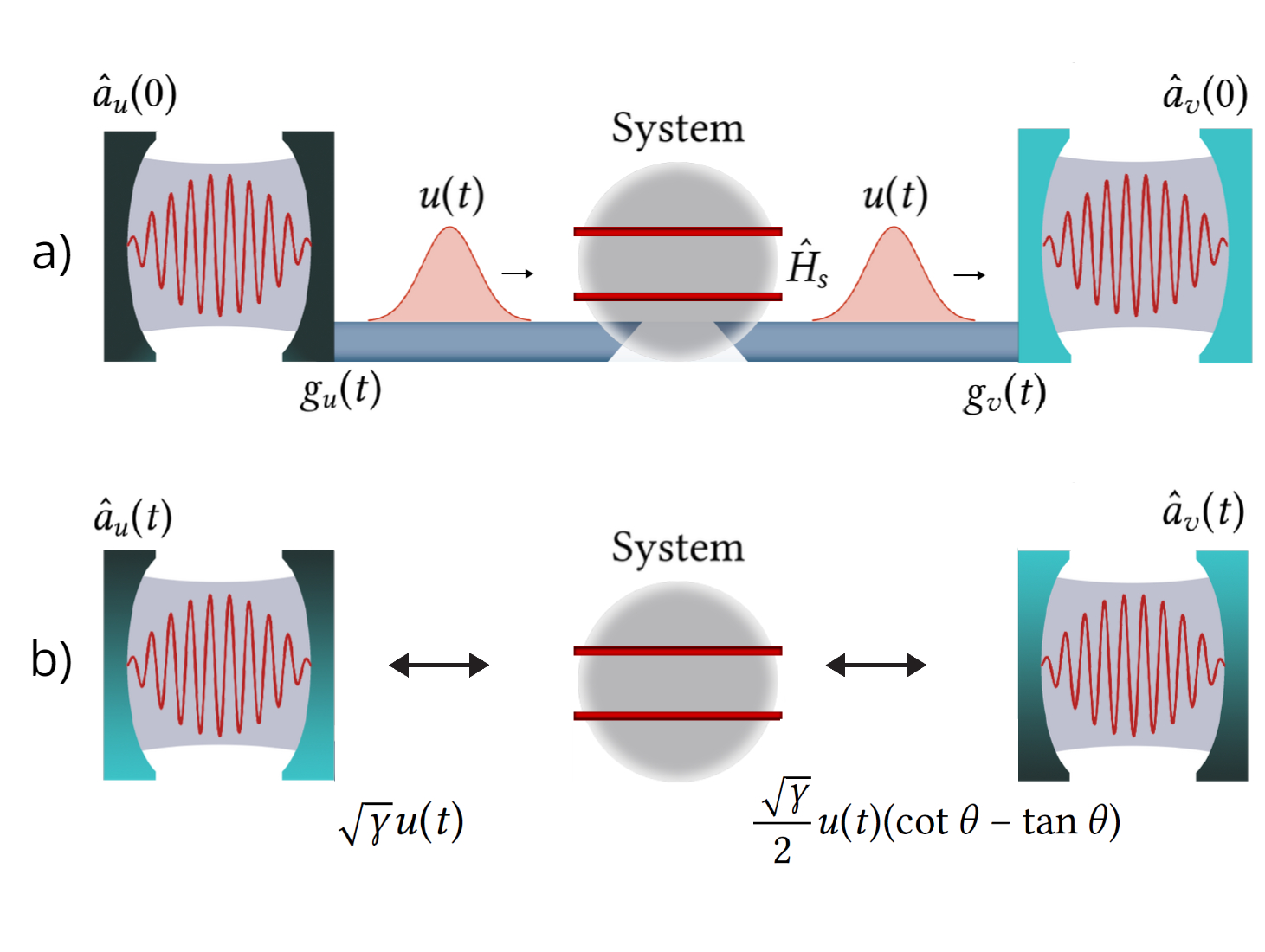}
    \caption{a) A pulse $u(t)$ is emitted from a virtual cavity with coupling $g_u(t)$ and interacts with a localized quantum system. Another virtual cavity with coupling $g_v(t)$ absorbs the quantum state contents of the outgoing pulse, chosen here of the same form $u(t)$. b) In the interaction picture, the scatterer interacts with two time dependent modes $a_u(t)$ and $a_v(t)$ with interaction strengths $\sqrt{\gamma}u(t)$ and $\frac{\sqrt{\gamma}}{2}u(t)(\cot\theta - \tan\theta)$. The  operators $\hat{a}_{u(v)}(t)$ act on time dependent superpositions of the incident and outgoing modes in panel (a), and they are graphically represented here by the left and right cavities.   $\hat{a}_{u(v)}(t)$ coincide initially with the Schr\"odinger picture mode operators $\hat{a}_{u(v)}(0)$ but with time they exchange character and become  $\hat{a}_{v(u)}(0)$.}
    \label{fig:rotated_system}
\end{figure}

A visual representation of the transformation to the interaction picture  is presented in Fig. \ref{fig:rotated_system}, where panel (a) shows the release and recapture of the pulse $u(t)$ by the virtual cavities with time dependent coupling strengths $\sqrt{\gamma} g_u(t)$ and $\sqrt{\gamma} g_v(t)$ to the scatterer in the Schr\"odinger picture. Panel (b) shows the coupling to the time evolved modes in the interaction picture, where the left most cavity represents the freely propagating pulse mode, cf., the coupling strength $\sqrt{\gamma} u(t)$,  and the right most cavity represents an orthogonal superposition of the two modes shown in panel (a). 

In absence of the coupling to the central system (setting $\gamma=0$) the Hamiltonian vanishes in the interaction picture, and nothing happens to the quantum content of the time dependent left and right modes in the interaction picture of Fig. \ref{fig:rotated_system}(b). When $\gamma\neq 0$, during the finite interaction time, there may be only a limited exchange of quanta between the scatterer and the time dependent modes which is in stark contrast with the complete emptying and partial filling of the $u$- and $v$-cavity modes in the Schr\"odinger picture represented by  Fig. \ref{fig:rotated_system}(a).  

\subsection{In what direction does the radiation propagate  in the interaction picture ?}
In our cascaded open system approach  \cite{short_kiilerich,long_kiilerich}, it is assumed, that the light only travels -- and the central system only emits -- in the direction from the initially populated $u$-cavity towards the output mode $v$-cavity.
\begin{figure}
    \centering
    \begin{minipage}{\columnwidth}
        \centering
        \includegraphics[width=\linewidth]{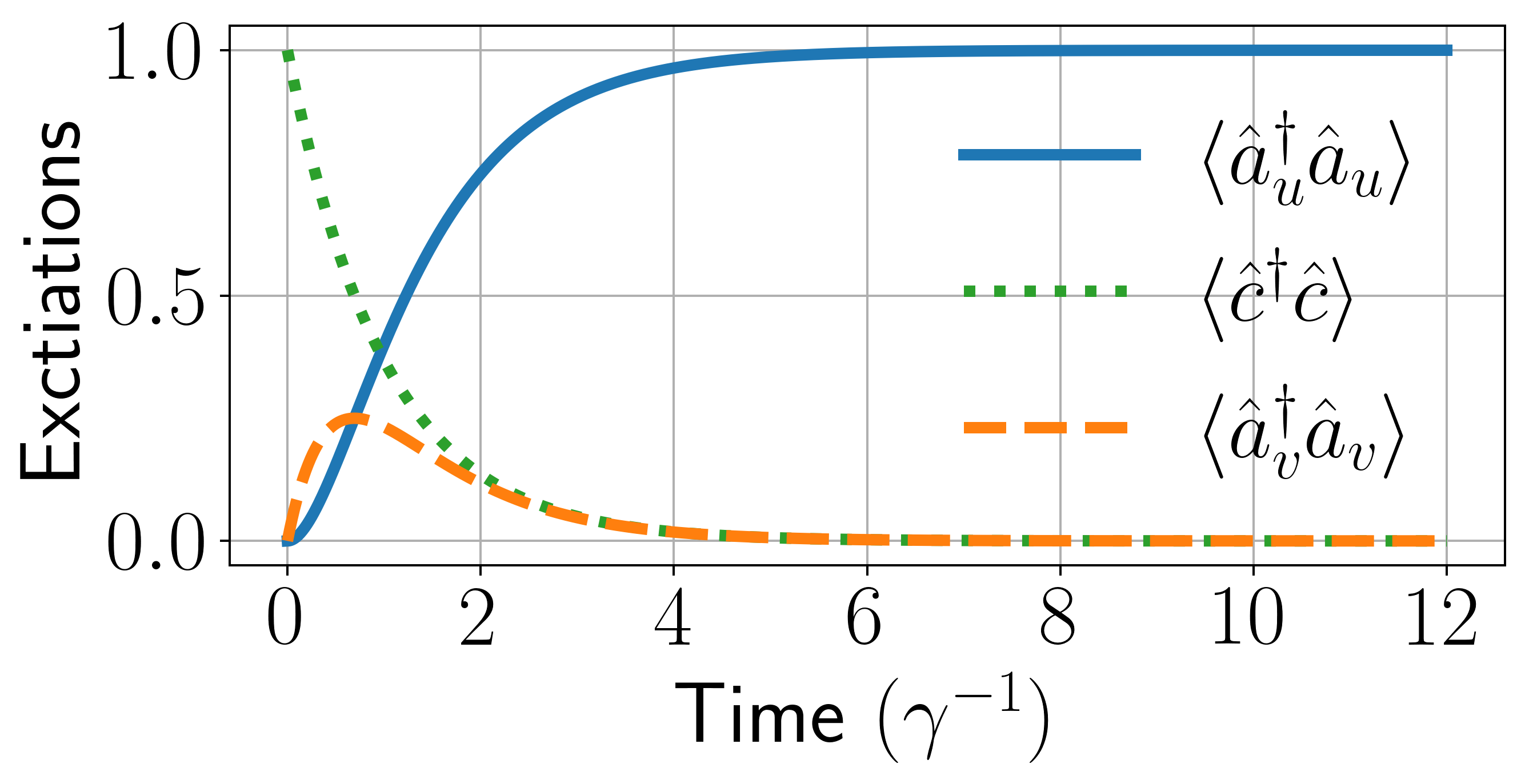}
        \caption{Decay of an excited two level system into a travelling pulse $u(t)$, treated here as an input vacuum pulse, and output single photon pulse. The interaction is described by a time dependent Jaynes-Cummings interaction and the coupling and correlated decay with an ancillary mode.}
        \label{fig:exponential_decay}
    \end{minipage}\hfill
    
\end{figure}
In the Schr\"odinger picture, terms in the Hamiltonian and the Lindblad damping terms thus exactly cancel  upstream propagation of excitations from the scatterer and the $v$-cavity. The same formal cancellation, however, is less obvious in the interaction picture where an initial excitation of the $v$-cavity seems to play a role in the dynamics due to the $\hat{c}^\dagger\hat{a}_v$  term in the Hamiltonian \eqref{eq:improved_interaction_hamiltonian}. A numerical calculation, however, shows that the $v$-cavity is instantly emptied, and to derive this result, we consider the  mean amplitude $\beta$ in the $v$-cavity mode, which decays by the rate
\begin{align}
    \frac{d\beta}{dt} = - \frac{|g_v|^2}{2} \beta.
\end{align}
Solving this equation by quadrature, we obtain
\begin{align}
    \int_{\beta_1}^{\beta_2} \frac{d\beta}{\beta} = - \frac{1}{2} \int_0^t dt' |g_v(t')|^2.
\end{align}
Now, we note  that $|g_v(t)|^2$ equals $|v(t)|^2$ divided by $F(t)=\int_0^t dt'|v(t')|^2$, and we define $f(t) = |v(t)|^2=dF/dt$ to obtain
\begin{align}
    \int_{\beta_1}^{\beta_2} \frac{d\beta}{\beta} = \ln{\frac{\beta_2}{\beta_1}} = - \frac{1}{2} \int_0^t dt' \frac{f(t')}{F(t')}.
\end{align}
The last term is readily evaluated and yields
\begin{align}
    \ln{\frac{\beta(t)}{\beta(0)}} =  - \frac{1}{2} \ln{\left( \frac{F(t)}{F(0)}\right)}
    \Rightarrow \beta(t) = \beta(0) \sqrt{\frac{F(0)}{F(t)}}.
\end{align}
Since $F(t)$ vanishes for $t=0$ and is finite for any $t>0$, any initial excitation in the $v$-cavity, indeed, decays instantly and there is no upstream propagation. 

\begin{figure}[h]
    \centering
    \includegraphics[width=\linewidth]{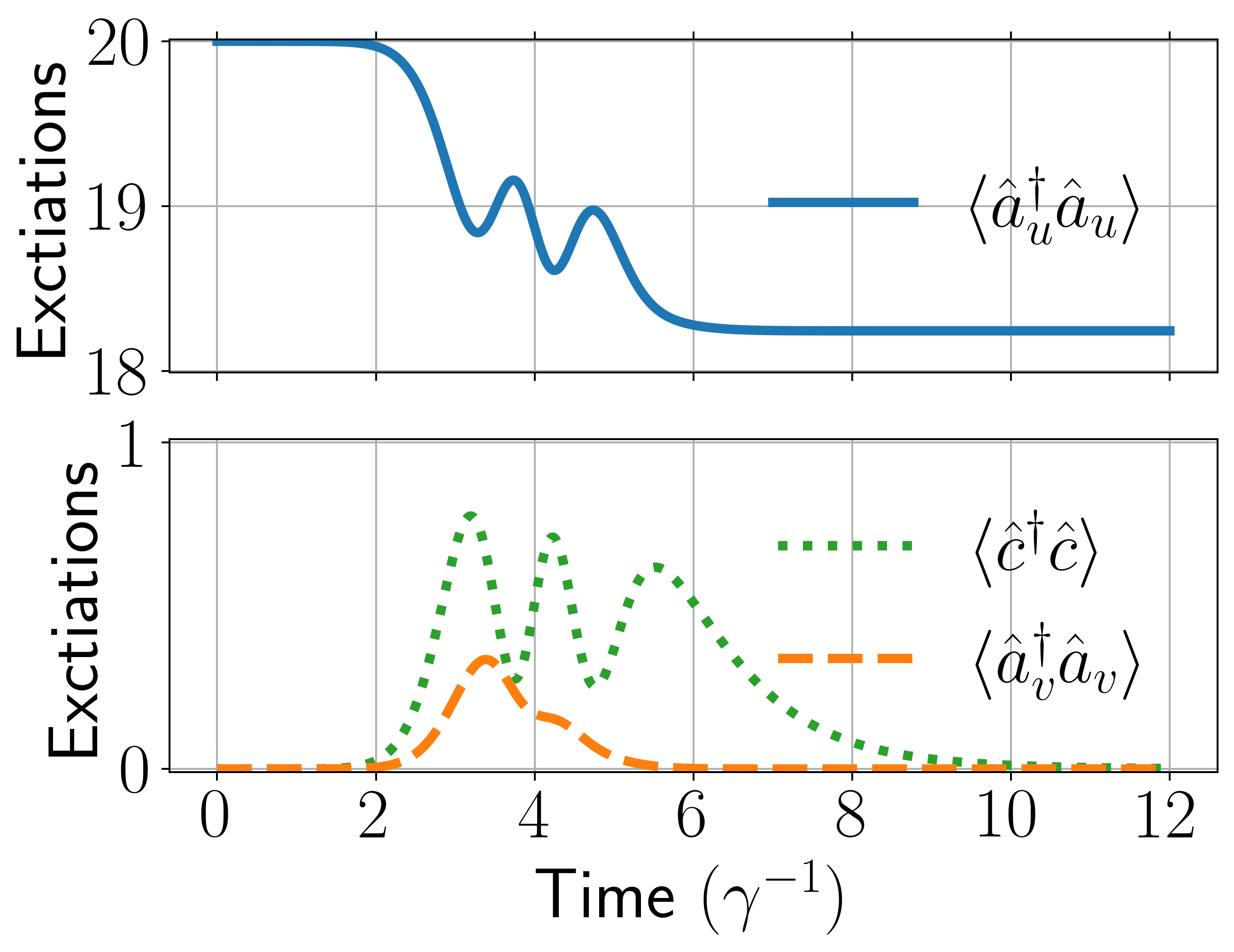}
    \caption{Simulation of Rabi oscillation dynamics due to excitation of a two-level system by an $n=20$ Fock state Gaussian pulse. Panel (a) shows how  the interaction picture $u$ oscillator is subject to the exchange of only few quanta of excitation with the scatterer and the $v$-oscillator. Panel (b) shows how the two level atom undergoes partial Rabi oscillations, out of phase with the mean excitation of the $u$-mode, confirming the exhange of quanta with the travelling pulse. The ancillary $v$-mode is temporarily excited while the net decrease of the total number of excitations in the final state is due to  incoherent loss to other modes and a possible coherent modification of the output pulse due to the interaction.}
    \label{fig:rabi_oscillations}
\end{figure}

\subsection{Rabi Oscillations with a quantum pulse} \label{sec:two_level_system}

Atoms excited by coherent  monochromatic light show  sinusoidal Rabi oscillations of their excited state population, $\rho_{22}(t) = \sin^2\left(\Omega t/2\right)$. In cavity QED, similar oscillations occur for an atom interacting with a quantized field  \cite{rabi_oscillations}, and we can now also investigate how an atom behaves when exposed to a resonant pulse $u(t)$ with quantum light.

We want to address the excitation of both the atom and of the travelling pulse, and we consider the output field mode $v(t) = u(t)$, described by the Hamiltonian \eqref{eq:improved_interaction_hamiltonian} and the Lindblad operator \eqref{eq:improved_lindblad}.  We use a normalized Gaussian pulse
\begin{align}
    u(t) = \frac{1}{\sqrt{\tau}\pi^{1/4}} \text{exp}\left(-\frac{(t-t_p)^2}{2\tau^2}\right),
\end{align}
with $t_p=4\gamma^{-1}$ and $\tau = 1\gamma^{-1}$ in the numerical calculations. With this choice Eq.\eqref{eq:theta_definition} yields
\begin{align}
    \theta(t) = \sin^{-1}\left( \sqrt{\frac{1}{2}\left(\text{erf}\left(\frac{t-t_p}{\tau}\right) + \text{erf}\left(\frac{t_p}{\tau}\right)\right)} \right),
\end{align}
where erf denotes the error function.

Fig. \ref{fig:rabi_oscillations} shows the time evolution, starting with the ground state atoms and an incident pulse prepared in a Fock state with $n=20$ photons. It is clear from the behavior of  $\langle \hat{c}^\dagger \hat{c} \rangle$ in the lower panel that the two-level system undergoes three Rabi oscillations before the interaction is over and the system relaxes to the ground state. 
In the Schr\"odinger picture calculation, the $u$-cavity mode is completely emptied of its initial excitation which is partially retrieved by the $v$-cavity mode, and the quantum state of the field hence explores the tensor product of two 21-dimensional Hilbert spaces, In the interaction picture, the time evolving oscillator modes explore only a few different Fock states. The upper panel in Fig.(\ref{fig:rabi_oscillations}) thus shows how the time dependent $u$-mode loses just 1-2 photons while the $v$-mode acquires less than unit excitation. Truncation of the Hilbert space to the relevant number state components thus easily yields a 20-fold reduction in dimension, and a corresponding 400-fold reduction in the number of density matrix elements. A factor that would be even larger for larger incident photon numbers and results in much shorter computing times.   

\section{Three-mode interaction picture} \label{sec:three_mode_interaction_picture}
In the previous section, we transformed to an interaction picture with respect to the coupling term between virtual $u$ and $v$, input and output cavity modes. If the scatterer is itself an optical cavity or a system contained inside a cavity, it is possible to apply an interaction picture with respect to the coupling terms among all these cavities. For a linear coupling of three cavities modes described by the Hamiltonian $\hat{H}_0$
\begin{widetext}
\begin{align}\label{eq:H0_3modes}
    \hat{H}_0(t) = \frac{i}{2}(\sqrt{\gamma}g_u(t)\hat{a}_u^\dagger\hat{c} 
    + \sqrt{\gamma}g_v^\ast(t)\hat{c}^\dagger\hat{a}_v
    + g_u(t)g_v^\ast(t)\hat{a}_u^\dagger\hat{a}_v - \text{H.c.}),
\end{align}
we define $\hat{U}_0(t)$ as the solution to Eq. \eqref{eq:U0}, and assume the ansatz
\begin{align}\label{eq:U0_3modes}
    \hat{U}_0(t) = \text{exp}(\lambda_1(t) \hat{a}_u^\dagger\hat{c} + \lambda_2(t) \hat{c}^\dagger\hat{a}_v + \lambda_3(t) \hat{a}_u^\dagger\hat{a}_v - \text{H.c.}).
\end{align}
From the Hamiltonian \eqref{eq:H0_3modes} and the Ansatz for the unitary evolution \eqref{eq:U0_3modes}, follows 
the coupled differential equations,
\begin{align}\label{eq:lambdas}
    \Dot{\lambda}_1(t) = \frac{1}{2}\sqrt{\gamma}g_u(t), \quad \Dot{\lambda}_2(t) = \frac{1}{2}\sqrt{\gamma}g_v^\ast(t), \quad \Dot{\lambda}_3(t) = \frac{1}{2}g_u(t)g_v^\ast(t).
\end{align}
\end{widetext}

We can expand the set of field operators in the interaction picture as a vector
\begin{align}\label{eq:phiM}
    \phi(t) = 
    \begin{pmatrix}
       \hat{a}_u(t) \\
       \hat{c}(t) \\
       \hat{a}_v(t)
    \end{pmatrix} = M(t)\phi(0),
\end{align}
where $\phi(0)$ is the vector of operators in the Schr\"odinger picture. 

The Ansatz for the interaction picture operators yields the linear system of equations
\begin{align}
    \frac{d}{dt}M(t) = F(t)M(t),
\end{align}
where the coefficient matrix $F(t)$ contains the derivatives of the $\lambda$-functions according to \eqref{eq:lambdas}
\begin{align}
    F(t) = \frac{1}{2}\begin{pmatrix}
       0 & \sqrt{\gamma}g_u(t) & g_u(t)g_v^\ast(t) \\
       -\sqrt{\gamma}g_u^\ast(t) & 0 & \sqrt{\gamma}g_v^\ast(t) \\
       -g_u^\ast(t)g_v(t) & -\sqrt{\gamma}g_v(t) & 0
    \end{pmatrix}.
\end{align}
This linear system of equation is readily solved and yields the transformation of any further components of the Hamiltonian to the interaction picture.

\subsection{Three-mode interaction picture with an empty cavity}
We assume three cavities with the ladder operators $\hat{a}_u$, $\hat{c}$ and $\hat{a}_v$ and a Gaussian $u$-pulse as specified in section \ref{sec:two_level_system}. In the absence of any further interactions, the transformation to the interaction picture handles all the dynamics, and no quanta ever leave the interaction picture  $\hat{a}_u$-mode while the interaction picture $\hat{c}$ and $\hat{a}_v$ are ``dark'' modes that never become populated during the interaction.

Note that the output mode after scattering on a single mode cavity with linewidth $\gamma$ and no internal losses is given in frequency space by $v(\omega) = r(\omega)u(\omega)$ \cite{short_kiilerich} where
\begin{align} \label{eq:cavt}
    r(\omega) = \frac{i(\omega-\omega_c) + \frac{\gamma}{2}}{i(\omega-\omega_c) - \frac{\gamma}{2}}.
\end{align}
At late times, the interaction picture  operator $\hat{a}_u$ accounts for all the incident photons which now populate the pulse mode transformed  according to \eqref{eq:cavt}.

\subsection{Three-mode interaction picture with a Kerr non-linear cavity: squeezing of a light pulse}

As a non-trivial example of the application of the three-mode interaction picture, we introduce a non-linear Kerr effect in the $c$-cavity,
\begin{align} \label{eq:cat_hamiltonian}
    H_s(t) = K(\hat{c}^\dagger(t)\hat{c}(t))^2,
\end{align}
where $K$ is a constant. The non-linear Kerr interaction $(\hat{c}^\dagger(t)\hat{c}(t))^2$ \cite{quantum_optics} acts on a coherent state by phase shifting each Fock state by an amount proportional to $n$. This effectively stretches the complex phase space distribution of the state and transforms the coherent state into a squeezed state. This picture readily applies to an intra-cavity field, but a long pulse incident on a cavity may at no time have all its photons inside the cavity and may thus not be subject to the full non-linear interaction, while a short pulse is spectrally broad and may reflect without even entering the cavity.

Our theory takes the spatial propagation properly into account, and our interaction picture restricts the evolution of the quantum states of the field to the number states initially occupied. We have found that using $K=0.02 \gamma$, this effect transforms a coherent state with $\alpha=4$ and the pulse parameters defined in section \ref{sec:two_level_system} into a squeezed state at the end of the interaction. States with larger coherent state amplitudes become more squeezed while larger values of the Kerr-interaction strength $K$ may distort the mode shape and cause deterioration of the single mode character and hence loss of squeezing. 

\begin{figure}
    \centering
    \includegraphics[width=\linewidth]{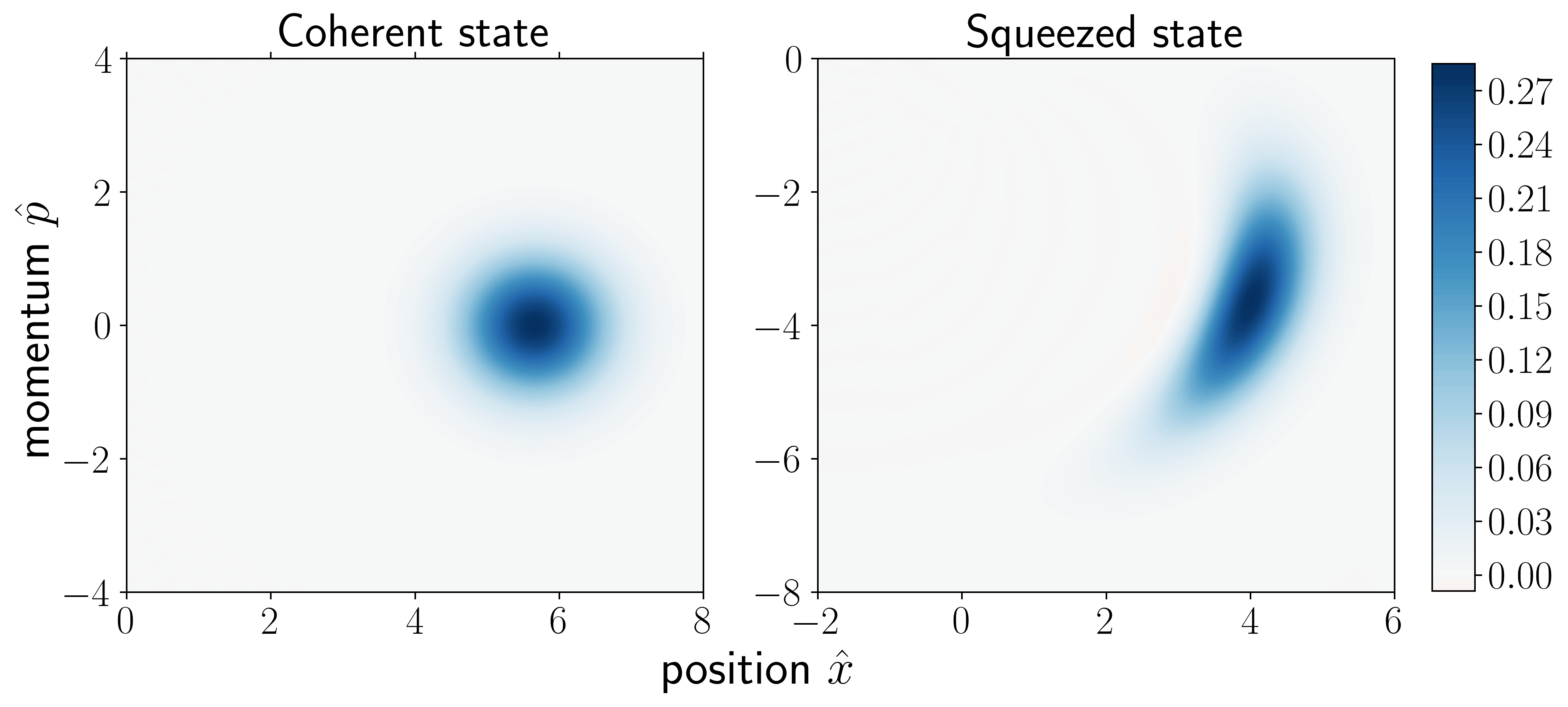}
    \caption{The Wigner function of an incident coherent state with $\alpha=4$ ($\langle x \rangle = 4\sqrt{2}$) (left panel) and the squeezed output state after passage of the pulse through a resonator with a Kerr-interaction with strength $K=0.02\gamma$ (right panel).}
    \label{fig:squeezed_state}
\end{figure}

\begin{figure}
    \centering
    \includegraphics[width=\linewidth]{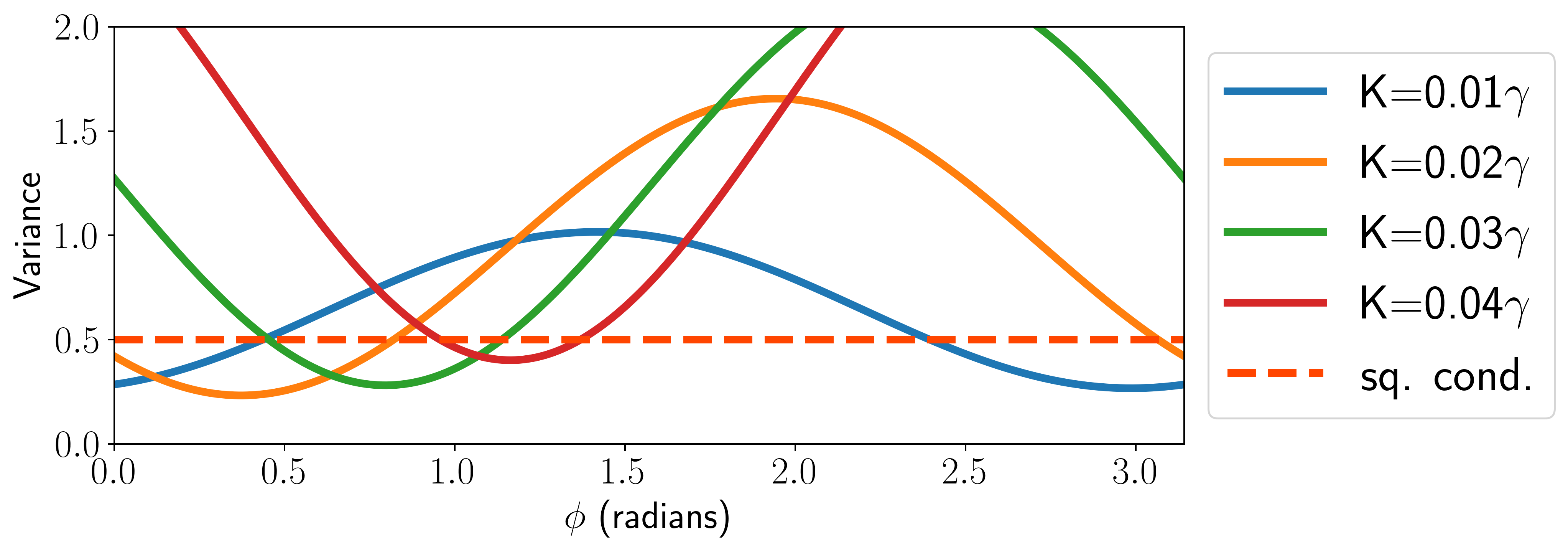}
    \caption{The variance of rotated field quadrature variables of a light pulse after passage of a non-linear resonator with different values of the Kerr non-linearity $K$. Values below the dashed line at the value 0.5 meet the the ``squeezing condition''.}
    \label{fig:squeezed_state_condition}
\end{figure}

We quantify the degree of squeezing by minimizing the variance of field quadratures rotated by the phase angle $\phi$,
\begin{align}
    \text{Var}(x_{\phi}) = \left\langle\left[\Delta (\cos{\phi}\hat{x} - \sin{\phi}\hat{p}) \right]^2 \right\rangle.
\end{align}
 For the case of an interaction with $K=0.02\gamma$ and $\alpha=4$, the minimum uncertainty arises for the angle $\phi = 0.52$ rad. where the variance is $\langle[\Delta (\cos{\phi}\hat{x} - \sin{\phi}\hat{p}) ]^2 \rangle \approx 0.245$ which is squeezed compared to the coherent state value of 0.5.  Fig.\ref{fig:squeezed_state} shows the Wigner function indeed resembles that of a squeezed state along an axis rotated clockwise by $\phi = 0.52 \; \mathrm{rad}\approx 30 \degree$. In Fig. \ref{fig:squeezed_state_condition}, the variance is shown as a function of $\phi$ for different values of the Kerr interaction strength, and we observe that the stronger Kerr effect leads to a smaller degree of (single mode) squeezing.
 

\subsection{Three-mode interaction picture with a Kerr non-linear cavity: turning a coherent pulse into a Schrödinger cat}

The value of $n^2$ can be written as $4m$ and $4m+1$ for even and odd $n$ respectively, where $m$ is an integer. For a sufficiently strong Kerr-interaction, the accumulated different phase factors attain the values $i$ on odd and $1$ on even Fock states  and hence transfer a coherent state in a single mode cavity into a Schrödinger cat state \cite{yurke-stoler}. 

\begin{figure}
    \centering
    \includegraphics[width=\linewidth]{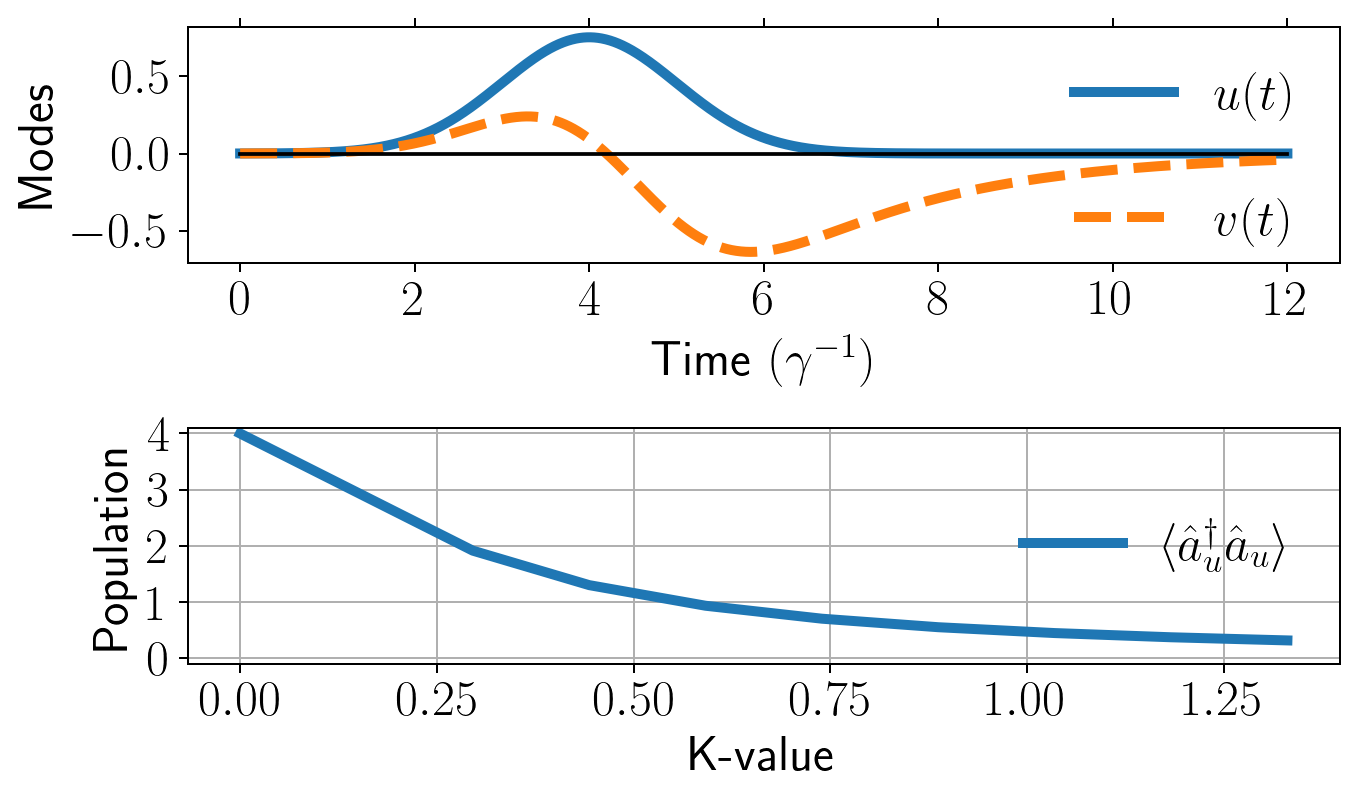}
    \caption{The upper panel shows the input $u(t)$ and output $v(t)$ modes for the scattering on an empty cavity. The lower panel shows the occupation of the output mode $v(t)$, represented asymptotically by $\hat{a}_u$ in the interaction picture, for different values of the non-linear interaction $K$. We assume an incident coherent state with $\langle \hat{n}\rangle = |\alpha|^2 = 4$, and no photons are lost in the process, but an increasing number of photons explore the continuum of modes orthogonal to $v(t)$.}
    \label{fig:one_interaction_cat_loss}
\end{figure}

\begin{figure}
    \centering
    \includegraphics[width=\linewidth]{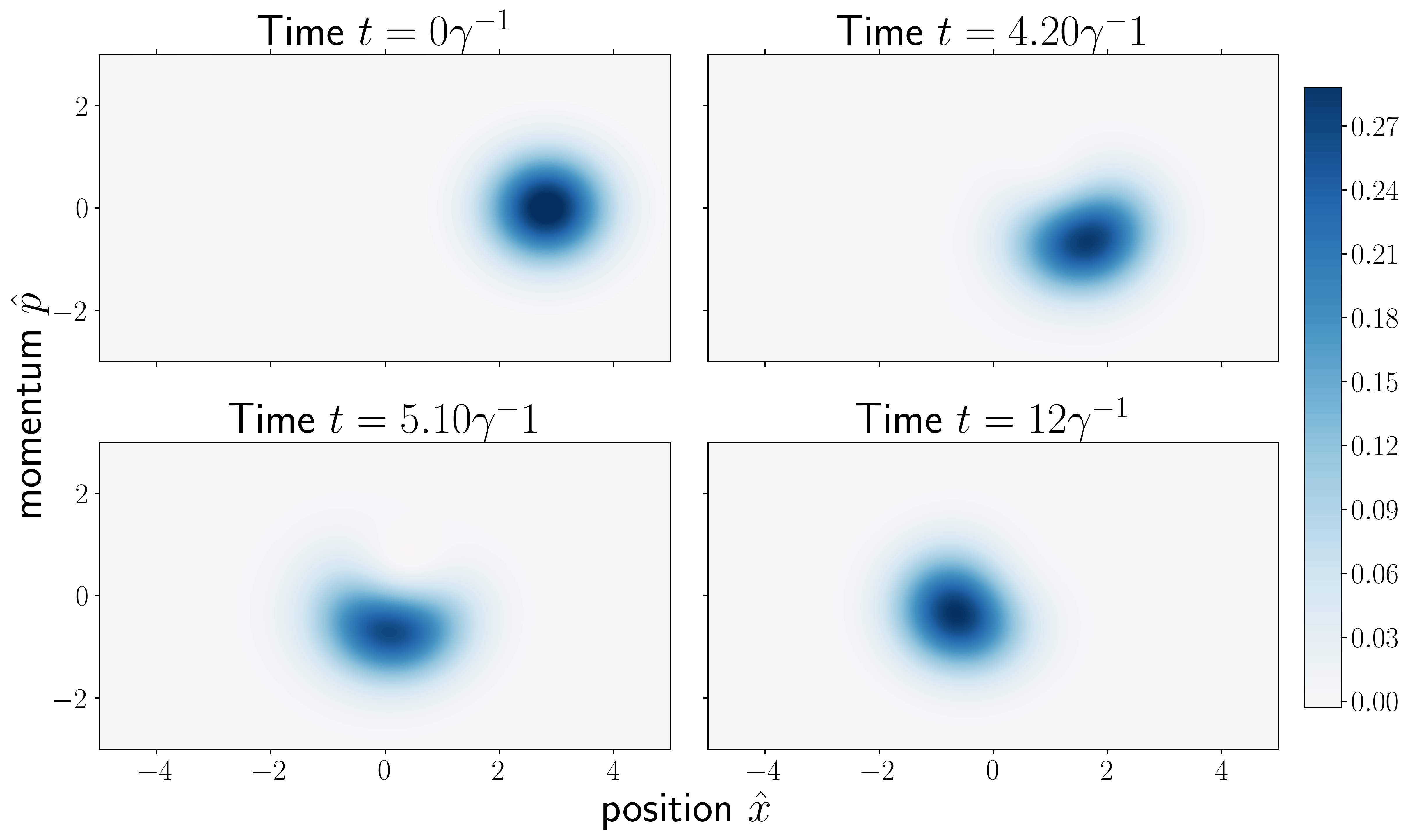}
    \caption{The Wigner function of the travelling pulse-mode at different times during the Kerr interaction, given by eq. \eqref{eq:cat_condition} for an initial coherent state with $\alpha=2$. It is clear that the interaction does not lead to a Schrödinger cat state.}
    \label{fig:one_interaction_cat}
\end{figure}

In our set-up, only the component of the pulse which is inside the non-linear $c$-mode cavity experiences the Kerr-interaction, which must be strong enough to yield the discrete phase differences to form the cat state, yet weak enough that it does not ruin the single mode character of the pulse. This poses a useful application of the interaction picture calculation, which focuses on the field content of the travelling pulse and takes the linear dispersion of the pulse shape by the passage of the middle cavity into account. In the absence of the Kerr interaction, the quantum state of the pulse is fully preserved and unchanged. Assuming the time-dependent operators in the interaction picture with respect to \eqref{eq:H0_3modes} and the expansion \eqref{eq:phiM}, for a weak Kerr interaction we may estimate its effect on the quantum state, exploiting the expression for the $\hat{c}(t)$-operator 
\begin{align}
    \hat{c}(t) = M_{21}(t)\hat{a}_u(0) + M_{22}(t)\hat{c}(0) + M_{23}(t)\hat{a}_v(0).
\end{align}
If we approximate $\hat{c}(t)$ by including only the contribution from the initially populated $\hat{a}_u$-mode, we have  $(\hat{c}^\dagger(t)\hat{c}(t))^2 \approx |M_{21}(t)|^4(\hat{a}_u^\dagger\hat{a}_u)^2 = |M_{21}(t)|^4 \hat{n}^2$, where the operator $\hat{n}^2$ is the square of the almost unchanged photon number in the pulse. The solution to Schrödinger's equation (in the interaction picture) for an initial coherent state is then
\begin{align}
    \ket{\psi(t)} = \text{exp}\left(-iK\hat{n}^2\int_0^tdt'|M_{21}(t')|^4\right) \ket{\alpha}.
\end{align}
To create the Yurke-Stoler state, the phase difference between even and odd $n$ should be $\pi/2$, and hence $K$ must satisfy
\begin{align} \label{eq:cat_condition}
    K = \frac{\pi}{2} \left(\int_0^T dt'|M_{21}(t')|^4 \right)^{-1},
\end{align}
where $T$ is the duration of the pulse. Assuming an incident Gaussian wave packet with the same parameters as in section \ref{sec:two_level_system}, the numerical evaluation of the integral over $M_{21}(t)$ yields a required value of $K\simeq 1.3$ to form the Schr\"odinger cat state. When we solve the three-mode problem numerically in the interaction picture with a sufficiently strong interaction, however, we find a significant loss of population of the $\hat{a}_u$ pulse modes, see Fig.\ref{fig:one_interaction_cat_loss}. Fig. \ref{fig:one_interaction_cat} shows that the Wigner function of the quantum state of the travelling pulse-mode at different times during the interaction. It is evident that the loss of amplitude also prevents the formation of the cat state.

It seems that the large value needed for $K$ is not compatible with the preservation of the single mode character of the quantum field. Instead of only looking at a single interaction, we therefore propose to let the light pulse pass the non-linear $c$-cavity several times (or pass through a sequence of several such cavities), and thus accumulate the non-linear phase shift from repeated weak Kerr-interactions with a small value of $K$. To ensure that the input pulse has  the same initial Gaussian shape at each interaction, one may need to reshape the pulse, e.g., by sum frequency generation \cite{silberhorn}. The weak Kerr interaction at each passage, transforms the quantum state of the field only  slightly, while maintaining its single mode character, and after sufficiently many passages, the light pulse attains the Schrödinger cat quantum state.

To create the cat by $N$ transmission steps,  Eq. \eqref{eq:cat_condition} can now be relaxed to
\begin{align}
    N\cdot K\int_0^Tdt'|M_{21}(t')|^4 = \frac{\pi}{2}.
\end{align}

\begin{figure}
    \centering
    \includegraphics[width=\linewidth]{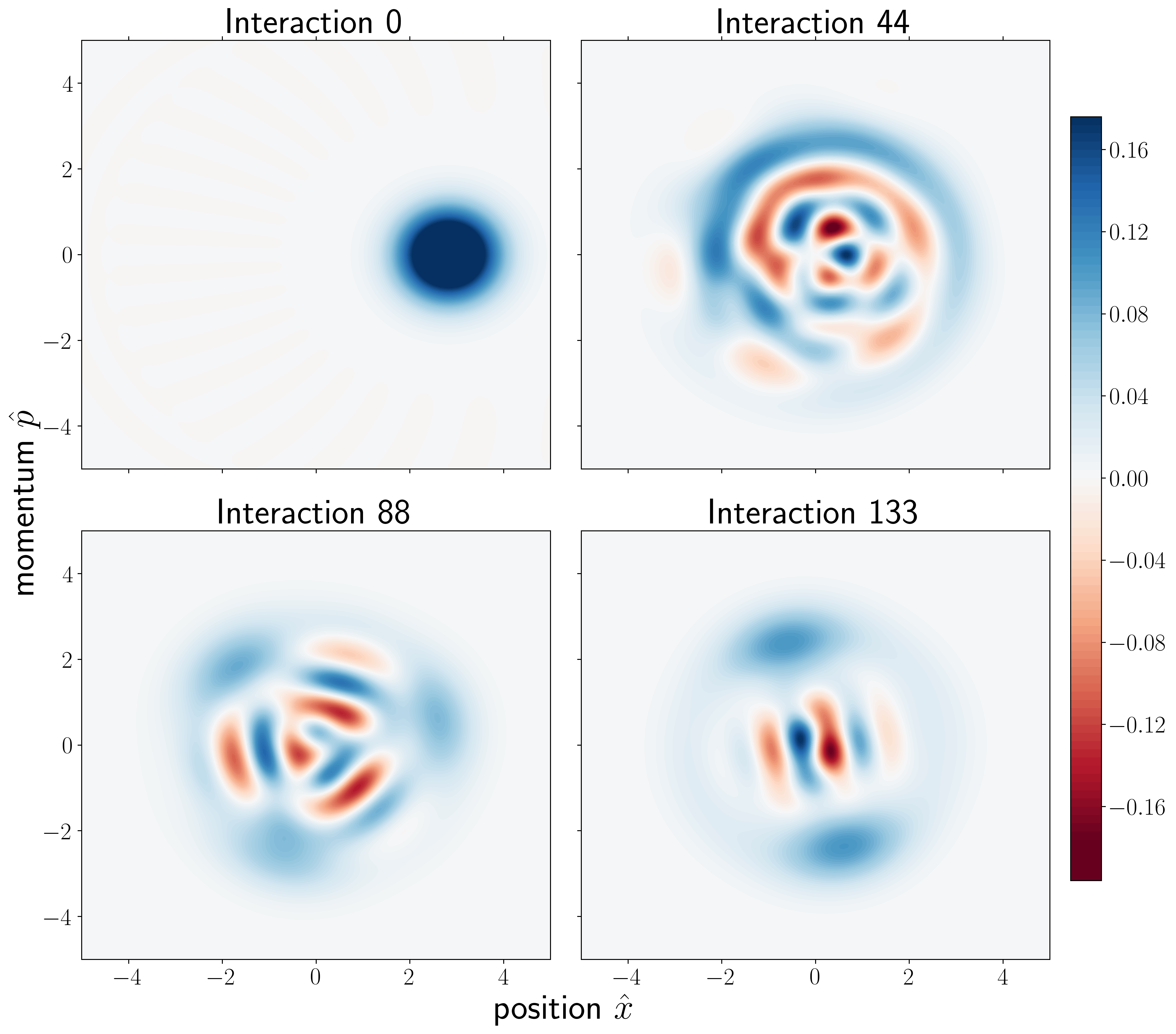}
    \caption{The Wigner function  of the quantum state of the transmitted pulse after different numbers of cavity transmissions. The simulations are carried out for $K=0.01\gamma$ and a coherent state with $\alpha=2$, and they assume that the output pulse of each cavity transmission is filtered and reshaped before transmission through the next cavity.}
    \label{fig:schrödinger_cat_state}
\end{figure}

Using an interaction strength $K=0.01\gamma$, a Gaussian wave packet with the same parameters as in section \ref{sec:two_level_system} and the numerical evaluation of the integral $\int_0^Tdt' |M_{21}(t')|\approx 1.180 \gamma^{-1}$, we estimate the number of interactions required to create a cat state from an input coherent state as
\begin{align}
    N =  \frac{\pi}{2 \cdot 0.01\gamma \cdot 1.180\gamma^{-1}} \approx 133.
\end{align}
This is in good agreement with the full numerical calculation, presented in Fig. \ref{fig:schrödinger_cat_state}, which shows the development of different non--classical states in the process towards the final Yurke-Stoler cat state. 

\section{Conclusion and outlook}

 An effective cascaded open quantum  system approach can be used to describe the initial preparation, the interaction and the final analysis of how a pulse of quantum radiation interacts with, e.g., an atom. The propagation of quantum pulses in free space and among cavities can be solved exactly in the Heisenberg picture by a linear transformation of the mode operators. This constitutes a good starting point for an interaction picture treatment of the cascaded master equation theory. The interaction between the propagating light and localized scatterers thus takes the conventional Jaynes-Cummings form of a time-dependent single mode interaction, while it also provides the interaction with an auxiliary oscillator. These terms together ensure the exact analysis assuming only the standard Born-Markov treatment of couplings to the continuum of free space (or guided) modes.

Our theory is simple to implement directly with standard master equation solvers, and since the main part of the dynamics is already taken care of by the transformation to the interaction picture, it considerably simplifies the numerical calculations. We note that our analysis may equally apply to other bosonic fields (light, microwaves, matter waves, sound and spin waves). For a wide range of wave phenomena, (linear) dispersion plays an important role and may be treated by a suitable interaction picture along the same lines as our treatment of the transmission through a single linear cavity. Our method thus permits dealing separately with the effects of propagation and of quantum interactions.

\section{Acknowledgements}

The authors acknowledge discussions with Fan Yang and Mads Middelhede Lund and support from the Euro-
pean Union FETFLAG program, Grant No. 820391
(SQUARE), the Danish National Research Foundation through the Center of Excellence for Complex Quantum Systems (Grant agreement No. DNRF156), and the European Union’s Horizon 2020 Research and Innovation Programme under the Marie Sklodowska-Curie program (754513).

\bibliography{contents/bibliography.bib}

\begin{thebibliography}{21}%
\makeatletter
\providecommand \@ifxundefined [1]{%
 \@ifx{#1\undefined}
}%
\providecommand \@ifnum [1]{%
 \ifnum #1\expandafter \@firstoftwo
 \else \expandafter \@secondoftwo
 \fi
}%
\providecommand \@ifx [1]{%
 \ifx #1\expandafter \@firstoftwo
 \else \expandafter \@secondoftwo
 \fi
}%
\providecommand \natexlab [1]{#1}%
\providecommand \enquote  [1]{``#1''}%
\providecommand \bibnamefont  [1]{#1}%
\providecommand \bibfnamefont [1]{#1}%
\providecommand \citenamefont [1]{#1}%
\providecommand \href@noop [0]{\@secondoftwo}%
\providecommand \href [0]{\begingroup \@sanitize@url \@href}%
\providecommand \@href[1]{\@@startlink{#1}\@@href}%
\providecommand \@@href[1]{\endgroup#1\@@endlink}%
\providecommand \@sanitize@url [0]{\catcode `\\12\catcode `\$12\catcode
  `\&12\catcode `\#12\catcode `\^12\catcode `\_12\catcode `\%12\relax}%
\providecommand \@@startlink[1]{}%
\providecommand \@@endlink[0]{}%
\providecommand \url  [0]{\begingroup\@sanitize@url \@url }%
\providecommand \@url [1]{\endgroup\@href {#1}{\urlprefix }}%
\providecommand \urlprefix  [0]{URL }%
\providecommand \Eprint [0]{\href }%
\providecommand \doibase [0]{http://dx.doi.org/}%
\providecommand \selectlanguage [0]{\@gobble}%
\providecommand \bibinfo  [0]{\@secondoftwo}%
\providecommand \bibfield  [0]{\@secondoftwo}%
\providecommand \translation [1]{[#1]}%
\providecommand \BibitemOpen [0]{}%
\providecommand \bibitemStop [0]{}%
\providecommand \bibitemNoStop [0]{.\EOS\space}%
\providecommand \EOS [0]{\spacefactor3000\relax}%
\providecommand \BibitemShut  [1]{\csname bibitem#1\endcsname}%
\let\auto@bib@innerbib\@empty
\bibitem [{\citenamefont {Kimble}(2008)}]{kimble2008quantum}%
  \BibitemOpen
  \bibfield  {author} {\bibinfo {author} {\bibfnamefont {H.~J.}\ \bibnamefont
  {Kimble}},\ }\bibfield  {title} {\enquote {\bibinfo {title} {The quantum
  internet},}\ }\href@noop {} {\bibfield  {journal} {\bibinfo  {journal}
  {Nature}\ }\textbf {\bibinfo {volume} {453}},\ \bibinfo {pages} {1023}
  (\bibinfo {year} {2008})}\BibitemShut {NoStop}%
\bibitem [{\citenamefont {Gheri}\ \emph {et~al.}(1998)\citenamefont {Gheri},
  \citenamefont {Ellinger}, \citenamefont {Pellizzari},\ and\ \citenamefont
  {Zoller}}]{gheri1998photon}%
  \BibitemOpen
  \bibfield  {author} {\bibinfo {author} {\bibfnamefont {K.~M.}\ \bibnamefont
  {Gheri}}, \bibinfo {author} {\bibfnamefont {K.}~\bibnamefont {Ellinger}},
  \bibinfo {author} {\bibfnamefont {T.}~\bibnamefont {Pellizzari}}, \ and\
  \bibinfo {author} {\bibfnamefont {P.}~\bibnamefont {Zoller}},\ }\bibfield
  {title} {\enquote {\bibinfo {title} {Photon-wavepackets as flying quantum
  bits},}\ }\href@noop {} {\bibfield  {journal} {\bibinfo  {journal}
  {Fortschritte der Physik: Progress of Physics}\ }\textbf {\bibinfo {volume}
  {46}},\ \bibinfo {pages} {401--415} (\bibinfo {year} {1998})}\BibitemShut
  {NoStop}%
\bibitem [{\citenamefont {Shen}\ and\ \citenamefont
  {Shen}(2015)}]{PhysRevA.92.033803}%
  \BibitemOpen
  \bibfield  {author} {\bibinfo {author} {\bibfnamefont {Y.}~\bibnamefont
  {Shen}}\ and\ \bibinfo {author} {\bibfnamefont {J.-T.}\ \bibnamefont
  {Shen}},\ }\bibfield  {title} {\enquote {\bibinfo {title}
  {{Photonic-Fock-state scattering in a waveguide-QED system and their
  correlation functions}},}\ }\href {\doibase 10.1103/PhysRevA.92.033803}
  {\bibfield  {journal} {\bibinfo  {journal} {Phys. Rev. A}\ }\textbf {\bibinfo
  {volume} {92}},\ \bibinfo {pages} {033803} (\bibinfo {year}
  {2015})}\BibitemShut {NoStop}%
\bibitem [{\citenamefont {Baragiola}\ \emph {et~al.}(2012)\citenamefont
  {Baragiola}, \citenamefont {Cook}, \citenamefont {Bra\ifmmode~\acute{n}\else
  \'{n}\fi{}czyk},\ and\ \citenamefont {Combes}}]{PhysRevA.86.013811}%
  \BibitemOpen
  \bibfield  {author} {\bibinfo {author} {\bibfnamefont {B.~Q.}\ \bibnamefont
  {Baragiola}}, \bibinfo {author} {\bibfnamefont {R.~L.}\ \bibnamefont {Cook}},
  \bibinfo {author} {\bibfnamefont {A.~M.}\ \bibnamefont
  {Bra\ifmmode~\acute{n}\else \'{n}\fi{}czyk}}, \ and\ \bibinfo {author}
  {\bibfnamefont {J.}~\bibnamefont {Combes}},\ }\bibfield  {title} {\enquote
  {\bibinfo {title} {$n$-photon wave packets interacting with an arbitrary
  quantum system},}\ }\href {\doibase 10.1103/PhysRevA.86.013811} {\bibfield
  {journal} {\bibinfo  {journal} {Phys. Rev. A}\ }\textbf {\bibinfo {volume}
  {86}},\ \bibinfo {pages} {013811} (\bibinfo {year} {2012})}\BibitemShut
  {NoStop}%
\bibitem [{\citenamefont {Gough}\ \emph {et~al.}(2012)\citenamefont {Gough},
  \citenamefont {James}, \citenamefont {Nurdin},\ and\ \citenamefont
  {Combes}}]{PhysRevA.86.043819}%
  \BibitemOpen
  \bibfield  {author} {\bibinfo {author} {\bibfnamefont {J.~E.}\ \bibnamefont
  {Gough}}, \bibinfo {author} {\bibfnamefont {M.~R.}\ \bibnamefont {James}},
  \bibinfo {author} {\bibfnamefont {H.~I.}\ \bibnamefont {Nurdin}}, \ and\
  \bibinfo {author} {\bibfnamefont {J.}~\bibnamefont {Combes}},\ }\bibfield
  {title} {\enquote {\bibinfo {title} {Quantum filtering for systems driven by
  fields in single-photon states or superposition of coherent states},}\ }\href
  {\doibase 10.1103/PhysRevA.86.043819} {\bibfield  {journal} {\bibinfo
  {journal} {Phys. Rev. A}\ }\textbf {\bibinfo {volume} {86}},\ \bibinfo
  {pages} {043819} (\bibinfo {year} {2012})}\BibitemShut {NoStop}%
\bibitem [{\citenamefont {Baragiola}\ and\ \citenamefont
  {Combes}(2017)}]{PhysRevA.96.023819}%
  \BibitemOpen
  \bibfield  {author} {\bibinfo {author} {\bibfnamefont {B.~Q.}\ \bibnamefont
  {Baragiola}}\ and\ \bibinfo {author} {\bibfnamefont {J.}~\bibnamefont
  {Combes}},\ }\bibfield  {title} {\enquote {\bibinfo {title} {Quantum
  trajectories for propagating fock states},}\ }\href {\doibase
  10.1103/PhysRevA.96.023819} {\bibfield  {journal} {\bibinfo  {journal} {Phys.
  Rev. A}\ }\textbf {\bibinfo {volume} {96}},\ \bibinfo {pages} {023819}
  (\bibinfo {year} {2017})}\BibitemShut {NoStop}%
\bibitem [{\citenamefont {Paris-Mandoki}\ \emph {et~al.}(2017)\citenamefont
  {Paris-Mandoki}, \citenamefont {Braun}, \citenamefont {Kumlin}, \citenamefont
  {Tresp}, \citenamefont {Mirgorodskiy}, \citenamefont {Christaller},
  \citenamefont {B\"uchler},\ and\ \citenamefont
  {Hofferberth}}]{PhysRevX.7.041010}%
  \BibitemOpen
  \bibfield  {author} {\bibinfo {author} {\bibfnamefont {A.}~\bibnamefont
  {Paris-Mandoki}}, \bibinfo {author} {\bibfnamefont {C.}~\bibnamefont
  {Braun}}, \bibinfo {author} {\bibfnamefont {J.}~\bibnamefont {Kumlin}},
  \bibinfo {author} {\bibfnamefont {C.}~\bibnamefont {Tresp}}, \bibinfo
  {author} {\bibfnamefont {I.}~\bibnamefont {Mirgorodskiy}}, \bibinfo {author}
  {\bibfnamefont {F.}~\bibnamefont {Christaller}}, \bibinfo {author}
  {\bibfnamefont {H.~P.}\ \bibnamefont {B\"uchler}}, \ and\ \bibinfo {author}
  {\bibfnamefont {S.}~\bibnamefont {Hofferberth}},\ }\bibfield  {title}
  {\enquote {\bibinfo {title} {Free-space quantum electrodynamics with a single
  {Rydberg} superatom},}\ }\href {\doibase 10.1103/PhysRevX.7.041010}
  {\bibfield  {journal} {\bibinfo  {journal} {Phys. Rev. X}\ }\textbf {\bibinfo
  {volume} {7}},\ \bibinfo {pages} {041010} (\bibinfo {year}
  {2017})}\BibitemShut {NoStop}%
\bibitem [{\citenamefont {ang Guofeng~Zhang}\ and\ \citenamefont
  {Wu}(2021)}]{zhang2021control}%
  \BibitemOpen
  \bibfield  {author} {\bibinfo {author} {\bibfnamefont {Wen-Long~Li}\
  \bibnamefont {ang Guofeng~Zhang}}\ and\ \bibinfo {author} {\bibfnamefont
  {Re-Bing}\ \bibnamefont {Wu}},\ }\href@noop {} {\enquote {\bibinfo {title}
  {On the control of flying qubits},}\ } (\bibinfo {year} {2021}),\ \Eprint
  {http://arxiv.org/abs/2111.00143} {arXiv:2111.00143 [quant-ph]} \BibitemShut
  {NoStop}%
\bibitem [{\citenamefont {Shen}\ and\ \citenamefont {Fan}(2005)}]{Shen:05}%
  \BibitemOpen
  \bibfield  {author} {\bibinfo {author} {\bibfnamefont {J.~T.}\ \bibnamefont
  {Shen}}\ and\ \bibinfo {author} {\bibfnamefont {Shanhui}\ \bibnamefont
  {Fan}},\ }\bibfield  {title} {\enquote {\bibinfo {title} {Coherent photon
  transport from spontaneous emission in one-dimensional waveguides},}\ }\href
  {\doibase 10.1364/OL.30.002001} {\bibfield  {journal} {\bibinfo  {journal}
  {Opt. Lett.}\ }\textbf {\bibinfo {volume} {30}},\ \bibinfo {pages}
  {2001--2003} (\bibinfo {year} {2005})}\BibitemShut {NoStop}%
\bibitem [{\citenamefont {Fischer}\ \emph {et~al.}(2018)\citenamefont
  {Fischer}, \citenamefont {Trivedi}, \citenamefont {Ramasesh}, \citenamefont
  {Siddiqi},\ and\ \citenamefont
  {Vu{\v{c}}kovi{\'{c}}}}]{Fischer2018scatteringintoone}%
  \BibitemOpen
  \bibfield  {author} {\bibinfo {author} {\bibfnamefont {Kevin~A.}\
  \bibnamefont {Fischer}}, \bibinfo {author} {\bibfnamefont {Rahul}\
  \bibnamefont {Trivedi}}, \bibinfo {author} {\bibfnamefont {Vinay}\
  \bibnamefont {Ramasesh}}, \bibinfo {author} {\bibfnamefont {Irfan}\
  \bibnamefont {Siddiqi}}, \ and\ \bibinfo {author} {\bibfnamefont {Jelena}\
  \bibnamefont {Vu{\v{c}}kovi{\'{c}}}},\ }\bibfield  {title} {\enquote
  {\bibinfo {title} {Scattering into one-dimensional waveguides from a
  coherently-driven quantum-optical system},}\ }\href {\doibase
  10.22331/q-2018-05-28-69} {\bibfield  {journal} {\bibinfo  {journal}
  {{Quantum}}\ }\textbf {\bibinfo {volume} {2}},\ \bibinfo {pages} {69}
  (\bibinfo {year} {2018})}\BibitemShut {NoStop}%
\bibitem [{\citenamefont {Shi}\ \emph {et~al.}(2015)\citenamefont {Shi},
  \citenamefont {Chang},\ and\ \citenamefont {Cirac}}]{shi2015}%
  \BibitemOpen
  \bibfield  {author} {\bibinfo {author} {\bibfnamefont {Tao}\ \bibnamefont
  {Shi}}, \bibinfo {author} {\bibfnamefont {Darrick~E.}\ \bibnamefont {Chang}},
  \ and\ \bibinfo {author} {\bibfnamefont {J.~Ignacio}\ \bibnamefont {Cirac}},\
  }\bibfield  {title} {\enquote {\bibinfo {title} {Multiphoton-scattering
  theory and generalized master equations},}\ }\href {\doibase
  10.1103/physreva.92.053834} {\bibfield  {journal} {\bibinfo  {journal}
  {Physical Review A}\ }\textbf {\bibinfo {volume} {92}} (\bibinfo {year}
  {2015}),\ 10.1103/physreva.92.053834}\BibitemShut {NoStop}%
\bibitem [{\citenamefont {Gardiner}\ and\ \citenamefont
  {Collett}(1985)}]{gardiner}%
  \BibitemOpen
  \bibfield  {author} {\bibinfo {author} {\bibfnamefont {C.~W.}\ \bibnamefont
  {Gardiner}}\ and\ \bibinfo {author} {\bibfnamefont {M.~J.}\ \bibnamefont
  {Collett}},\ }\bibfield  {title} {\enquote {\bibinfo {title} {Input and
  output in damped quantum systems: Quantum stochastic differential equations
  and the master equation},}\ }\href {\doibase 10.1103/PhysRevA.31.3761}
  {\bibfield  {journal} {\bibinfo  {journal} {Phys. Rev. A}\ }\textbf {\bibinfo
  {volume} {31}},\ \bibinfo {pages} {3761--3774} (\bibinfo {year}
  {1985})}\BibitemShut {NoStop}%
\bibitem [{\citenamefont {Carmichael}(1993)}]{Carmichael1993}%
  \BibitemOpen
  \bibfield  {author} {\bibinfo {author} {\bibfnamefont {H.~J.}\ \bibnamefont
  {Carmichael}},\ }\bibfield  {title} {\enquote {\bibinfo {title} {Quantum
  trajectory theory for cascaded open systems},}\ }\href {\doibase
  10.1103/PhysRevLett.70.2273} {\bibfield  {journal} {\bibinfo  {journal}
  {Phys. Rev. Lett.}\ }\textbf {\bibinfo {volume} {70}},\ \bibinfo {pages}
  {2273--2276} (\bibinfo {year} {1993})}\BibitemShut {NoStop}%
\bibitem [{\citenamefont {Kiilerich}\ and\ \citenamefont
  {M\o{}lmer}(2019)}]{short_kiilerich}%
  \BibitemOpen
  \bibfield  {author} {\bibinfo {author} {\bibfnamefont {Alexander~Holm}\
  \bibnamefont {Kiilerich}}\ and\ \bibinfo {author} {\bibfnamefont {Klaus}\
  \bibnamefont {M\o{}lmer}},\ }\bibfield  {title} {\enquote {\bibinfo {title}
  {Input-output theory with quantum pulses},}\ }\href {\doibase
  10.1103/PhysRevLett.123.123604} {\bibfield  {journal} {\bibinfo  {journal}
  {Phys. Rev. Lett.}\ }\textbf {\bibinfo {volume} {123}},\ \bibinfo {pages}
  {123604} (\bibinfo {year} {2019})}\BibitemShut {NoStop}%
\bibitem [{\citenamefont {Kiilerich}\ and\ \citenamefont
  {M\o{}lmer}(2020)}]{long_kiilerich}%
  \BibitemOpen
  \bibfield  {author} {\bibinfo {author} {\bibfnamefont {Alexander~Holm}\
  \bibnamefont {Kiilerich}}\ and\ \bibinfo {author} {\bibfnamefont {Klaus}\
  \bibnamefont {M\o{}lmer}},\ }\bibfield  {title} {\enquote {\bibinfo {title}
  {Quantum interactions with pulses of radiation},}\ }\href {\doibase
  10.1103/PhysRevA.102.023717} {\bibfield  {journal} {\bibinfo  {journal}
  {Phys. Rev. A}\ }\textbf {\bibinfo {volume} {102}},\ \bibinfo {pages}
  {023717} (\bibinfo {year} {2020})}\BibitemShut {NoStop}%
\bibitem [{\citenamefont {Lodahl}\ \emph {et~al.}(2017)\citenamefont {Lodahl},
  \citenamefont {Mahmoodian}, \citenamefont {Stobbe}, \citenamefont
  {Rauschenbeutel}, \citenamefont {Schneeweiss}, \citenamefont {Volz},
  \citenamefont {Pichler},\ and\ \citenamefont {Zoller}}]{lodahl2017chiral}%
  \BibitemOpen
  \bibfield  {author} {\bibinfo {author} {\bibfnamefont {P.}~\bibnamefont
  {Lodahl}}, \bibinfo {author} {\bibfnamefont {S.}~\bibnamefont {Mahmoodian}},
  \bibinfo {author} {\bibfnamefont {S.}~\bibnamefont {Stobbe}}, \bibinfo
  {author} {\bibfnamefont {A.}~\bibnamefont {Rauschenbeutel}}, \bibinfo
  {author} {\bibfnamefont {P.}~\bibnamefont {Schneeweiss}}, \bibinfo {author}
  {\bibfnamefont {J.}~\bibnamefont {Volz}}, \bibinfo {author} {\bibfnamefont
  {H.}~\bibnamefont {Pichler}}, \ and\ \bibinfo {author} {\bibfnamefont
  {P.}~\bibnamefont {Zoller}},\ }\bibfield  {title} {\enquote {\bibinfo {title}
  {Chiral quantum optics},}\ }\href@noop {} {\bibfield  {journal} {\bibinfo
  {journal} {Nature}\ }\textbf {\bibinfo {volume} {541}},\ \bibinfo {pages}
  {473} (\bibinfo {year} {2017})}\BibitemShut {NoStop}%
\bibitem [{\citenamefont {Pichler}\ and\ \citenamefont
  {Zoller}(2016)}]{zoller}%
  \BibitemOpen
  \bibfield  {author} {\bibinfo {author} {\bibfnamefont {Hannes}\ \bibnamefont
  {Pichler}}\ and\ \bibinfo {author} {\bibfnamefont {Peter}\ \bibnamefont
  {Zoller}},\ }\bibfield  {title} {\enquote {\bibinfo {title} {Photonic
  circuits with time delays and quantum feedback},}\ }\href {\doibase
  10.1103/PhysRevLett.116.093601} {\bibfield  {journal} {\bibinfo  {journal}
  {Phys. Rev. Lett.}\ }\textbf {\bibinfo {volume} {116}},\ \bibinfo {pages}
  {093601} (\bibinfo {year} {2016})}\BibitemShut {NoStop}%
\bibitem [{\citenamefont {Agarwal}(1985)}]{rabi_oscillations}%
  \BibitemOpen
  \bibfield  {author} {\bibinfo {author} {\bibfnamefont {G.~S.}\ \bibnamefont
  {Agarwal}},\ }\bibfield  {title} {\enquote {\bibinfo {title} {Vacuum-field
  rabi oscillations of atoms in a cavity},}\ }\href {\doibase
  10.1364/JOSAB.2.000480} {\bibfield  {journal} {\bibinfo  {journal} {J. Opt.
  Soc. Am. B}\ }\textbf {\bibinfo {volume} {2}},\ \bibinfo {pages} {480--485}
  (\bibinfo {year} {1985})}\BibitemShut {NoStop}%
\bibitem [{\citenamefont {{Gerry}}\ and\ \citenamefont
  {P.}(2005)}]{quantum_optics}%
  \BibitemOpen
  \bibfield  {author} {\bibinfo {author} {\bibfnamefont {C.}~\bibnamefont
  {{Gerry}}}\ and\ \bibinfo {author} {\bibfnamefont {{Knight}}\ \bibnamefont
  {P.}},\ }\enquote {\bibinfo {title} {{Introductory Quantum Optics}},}\ \
  (\bibinfo  {publisher} {Cambridge University Press, New York},\ \bibinfo
  {year} {2005})\ \bibinfo {edition} {1st}\ ed.\BibitemShut {Stop}%
\bibitem [{\citenamefont {Yurke}\ and\ \citenamefont
  {Stoler}(1986)}]{yurke-stoler}%
  \BibitemOpen
  \bibfield  {author} {\bibinfo {author} {\bibfnamefont {B.}~\bibnamefont
  {Yurke}}\ and\ \bibinfo {author} {\bibfnamefont {D.}~\bibnamefont {Stoler}},\
  }\bibfield  {title} {\enquote {\bibinfo {title} {Generating quantum
  mechanical superpositions of macroscopically distinguishable states via
  amplitude dispersion},}\ }\href {\doibase 10.1103/PhysRevLett.57.13}
  {\bibfield  {journal} {\bibinfo  {journal} {Phys. Rev. Lett.}\ }\textbf
  {\bibinfo {volume} {57}},\ \bibinfo {pages} {13--16} (\bibinfo {year}
  {1986})}\BibitemShut {NoStop}%
\bibitem [{\citenamefont {Eckstein}\ \emph {et~al.}(2011)\citenamefont
  {Eckstein}, \citenamefont {Brecht},\ and\ \citenamefont
  {Silberhorn}}]{silberhorn}%
  \BibitemOpen
  \bibfield  {author} {\bibinfo {author} {\bibfnamefont {Andreas}\ \bibnamefont
  {Eckstein}}, \bibinfo {author} {\bibfnamefont {Benjamin}\ \bibnamefont
  {Brecht}}, \ and\ \bibinfo {author} {\bibfnamefont {Christine}\ \bibnamefont
  {Silberhorn}},\ }\bibfield  {title} {\enquote {\bibinfo {title} {A quantum
  pulse gate based on spectrally engineered sum frequency generation},}\ }\href
  {\doibase 10.1364/OE.19.013770} {\bibfield  {journal} {\bibinfo  {journal}
  {Opt. Express}\ }\textbf {\bibinfo {volume} {19}},\ \bibinfo {pages}
  {13770--13778} (\bibinfo {year} {2011})}\BibitemShut {NoStop}%
\end{thebibliography}%

\end{document}